\documentclass[aps,apl,reprint,showpacs,superscriptaddress,twocolumn]{revtex4}
\usepackage{graphicx}
\usepackage{amsmath}
\usepackage{siunitx}
\usepackage{booktabs}

\begin{document}

\title{Thermoelastic study of nanolayered structures using time-resolved x-ray diffraction at high repetition rate}

\author{H.~Navirian}
\author{D.~Schick}
\email{daniel.schick@uni-potsdam.de}
\affiliation{Institut für Physik und Astronomie, Universität Potsdam, Karl-Liebknecht-Str. 24-25, 14476 Potsdam, Germany}
\author{P.~Gaal}
\affiliation{Helmholtz-Zentrum Berlin für Materialien und Energie GmbH, Wilhelm-Conrad-Röntgen Campus, BESSY II, Albert-Einstein-Str. 15, 12489 Berlin Germany}
\author{W.~Leitenberger}
\affiliation{Institut für Physik und Astronomie, Universität Potsdam, Karl-Liebknecht-Str. 24-25, 14476 Potsdam, Germany}
\author{R.~Shayduk}
\affiliation{Helmholtz-Zentrum Berlin für Materialien und Energie GmbH, Wilhelm-Conrad-Röntgen Campus, BESSY II, Albert-Einstein-Str. 15, 12489 Berlin Germany}
\author{M.~Bargheer}
\affiliation{Institut für Physik und Astronomie, Universität Potsdam, Karl-Liebknecht-Str. 24-25, 14476 Potsdam, Germany}

\date{\today}

\begin{abstract}
We investigate the thermoelastic response of a nanolayered sample composed of a metallic SrRuO$_3$ (SRO) electrode sandwiched between a ferroelectric Pb(Zr$_{0.2}$Ti$_{0.8}$)O$_3$ (PZT) film with negative thermal expansion and a SrTiO$_3$ substrate. SRO is rapidly heated by fs-laser pulses with \SI{208}{kHz} repetition rate. Diffraction of x-ray pulses derived from a synchrotron measures the transient out-of-plane lattice constant $c$ of all three materials simultaneously from \SI{120}{ps} to 5 $\mu$s with a relative accuracy up to $\Delta c/c=10^{-6}$. 
The in-plane propagation of sound is essential for understanding the delayed out of plane expansion.
\end{abstract}


\maketitle

\textbf{Introduction}

Ultrafast heat generation and transport in nanostructures are challenging the classical models of thermoelasticity.
For nanosized structures, becoming as small as the heat carrying phonon mean free path, Fourier theory of thermal transport can differ dramatically from bulk samples.\cite{good1999a,cahi2003a,siem2010a,luck2012a} The material properties on the nanoscale may also differ from equilibrium values for bulk material due to size effects or growth dependent interface resistance.
Quasi-instantaneous heating, e.g. by ultrashort laser pulses may lead to non-equilibrium phonon distributions and is much faster than thermal expansion, which is limited by the speed of sound in the material.
The resulting dynamic thermal expansion\cite{tang1991a,lee2009a,stou2012a} due to impulsive thermal stresses has been studied extensively in the context of coherent phonon excitation.\cite{thom1986a,wrig1992a,rose1999a,tang2008a}


Besides the general physical interest, the knowledge of the structural response to instantaneous thermal stresses becomes important for the performance and reliability of novel applications in nanostructures such as optoelectronic devices, MEMS, SASER, and X-ray optics for free electron lasers, in which heat is generated on ultrafast timescales.\cite{good1999a,cahi2003a,fain2013a,shvy2010a}

Numerical models for the calculation of the complete thermoelastic dynamics on the atomic level exist.\cite{vere2013a}
They require extensive computational power and a large number of material specific thermal and mechanical constants for an anisotropic sample geometry. Time-resolved experiments, which can access the material-specific structural information on the relevant length and time scales are an essential experimental cross-check for the complete understanding of the complex thermoelastic dynamics in flash-heated nanostructures.

In this study we applied ultrafast X-ray diffraction (UXRD) to measure the transient strains in three layers of a thin film heterostructure of functional oxide materials simultaneously by exploiting the material specific Bragg reflections. The large penetration depth of hard X-rays yields information also on the buried layers. 
The combination of ultrashort X-ray pulses with ultrafast Time-Correlated-Single-Photon-Counting (TCSPC) enables covering a pump-probe delay range of nearly five decades and ultra-sensitive detection of lattice constant changes down to $\Delta c/c = 10^{-6}$.
Such precise and continuous experimental information provide an excellent experimental test ground for simulations which attempt the description of the physics involving the coupling of heat and strain in complex materials on several time- and lengths scales.

We focus on the experimental strategy and confirm the interpretation by comparing experimental data to a simplified numerical model\cite{shay2011a}, which treats heat transport and thermal expansion sequentially and briefly discuss the deviations which show the complex three dimensional nature of the problem. Although the thin film structure suggests that a reduced 1D model should be appropriate for short timescales, the comparison of experiment and simulation reveals more complex thermoelastic dynamics from the picosecond to microsecond time scales.

\textbf{Experimental setup}

The experiments were performed on a typical ferroelectric thin film structure. A ferroelectric Pb(Zr$_{0.2}$Ti$_{0.8}$)O$_3$ (PZT) layer and a metallic SrRuO$_3$ (SRO) transducer layer were grown onto a SrTiO$_3$ (STO) substrate by pulsed laser deposition (PLD).\cite{vrej2006b}
From the detailed characterization by transmission electron microscopy (TEM) and static X-ray diffraction (XRD) in Fig. 1, we derived the layer ticknesses of $d_\text{PZT} = 207$~nm and $d_\text{SRO} = 147$~nm, as well as the average lattice constants normal to the sample surface of $c_\text{PZT} = 4.130$~\AA, $c_\text{SRO} = 3.948$~\AA, and $c_\text{STO} = 3.905$~\AA.\cite{schi2013a}

\begin{figure}
\center{
\includegraphics[width=0.8\columnwidth]{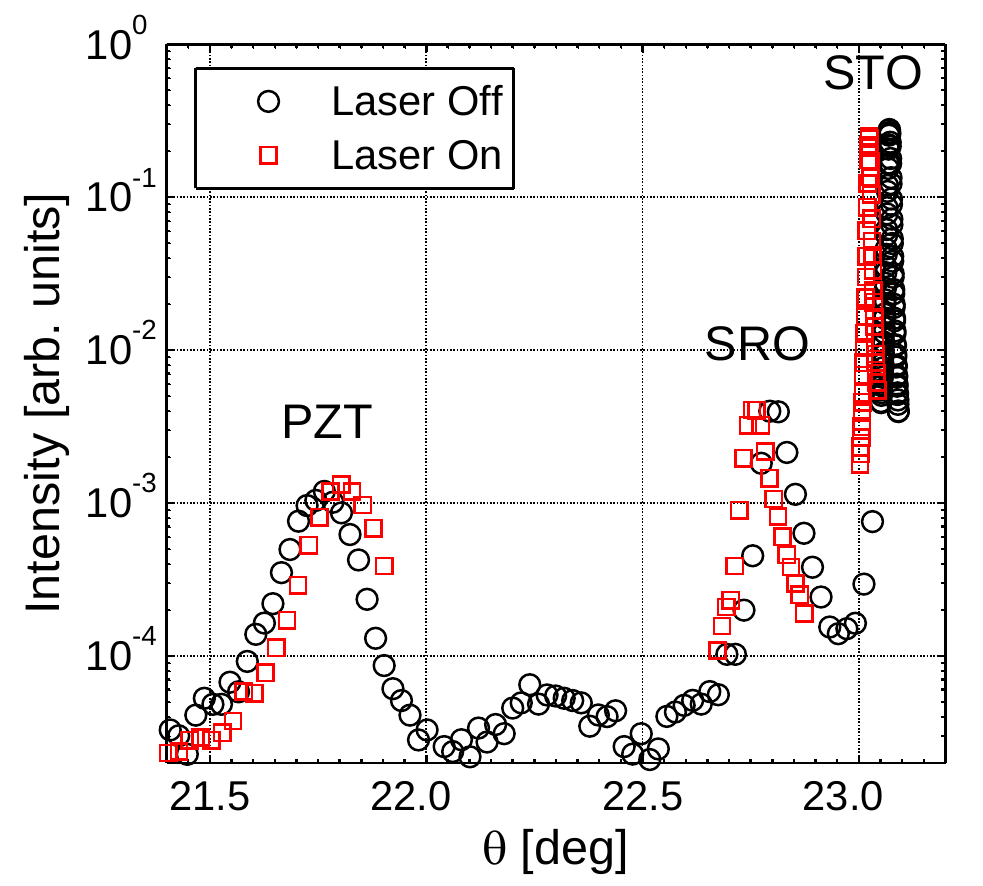}}
\caption{(Color online) Static rocking curve of PZT, SRO, and STO (0\,0\,2) Bragg reflections (black circles) with no laser on the sample. The red squares show the peak shift of the individual Bragg peaks due to an average temperature increase of approx. \SI{191}{K} for the laser-on state.}
\label{fig:rocking}
\end{figure}
The UXRD experiments were carried out at the EDR bending magnet beamline at BESSY II at Helmholtz Zentrum in Berlin.
The details of the setup were described recently.\cite{navi2012a} 
In order to probe a homogeneously excited sample area with 8.9~keV X-rays with a bandwidth of $\delta E/E = 2\times10^{-4}$ the beam is focused in the dimension perpendicular to the plane of diffraction and collimated in the diffraction plane down to a spot size of $50\times\SI{50}{\micro\metre\squared}$.
The symmetrically diffracted photons were collected by a point-detector with a plastic scintillator ($<$ 0.5\,ns rise time).
For pumping the sample an Ytterbium-doped fiber oscillator and a three-stage-amplifier laser system were electronically synchronized to the master clock of the storage ring. The laser repetition rate was set to 208 kHz thus determining the pump-probe delay range to 4.8 $\mu$s.\cite{navi2012a}.
The pump pulses with maximum pulse energy  $E_\text{P} = \SI{10}{\micro\joule}$ at a wavelength $\lambda = \SI{1030}{nm}$ and pulse duration of $\tau = \SI{250}{fs}$, were sent to the sample via a mechanical delay line.

\textbf{Results \& Discussion}

The transient thermoelastic response of the sample is measured by scanning the material-specific (0\,0\,2) Bragg reflections for various pump-probe delays.
The static rocking curve is shown in Fig.~\ref{fig:rocking} as black circles.
Although the sample is excited with femtosecond laser pulses, the static heat load due to the average pump power
leads to a stationary heating and corresponding thermal strains according to the individual expansion coefficients
observable by the shifted Bragg peaks (red squares) in Fig.~\ref{fig:rocking}. 
We use the literature value of the linear thermal expansion coefficients for STO and SRO (see Tab.~\ref{tab:parameter}) to deduce the corresponding temperature rise of the whole sample by approx. $\Delta T=190$K, starting from room temperature. For the transient experiments we used nearly two times smaller average laser power causing a smaller temperature rise of approx. $\Delta T=84$K (data not shown).
\begin{figure}
	\centering
	\includegraphics [width=0.8\columnwidth]{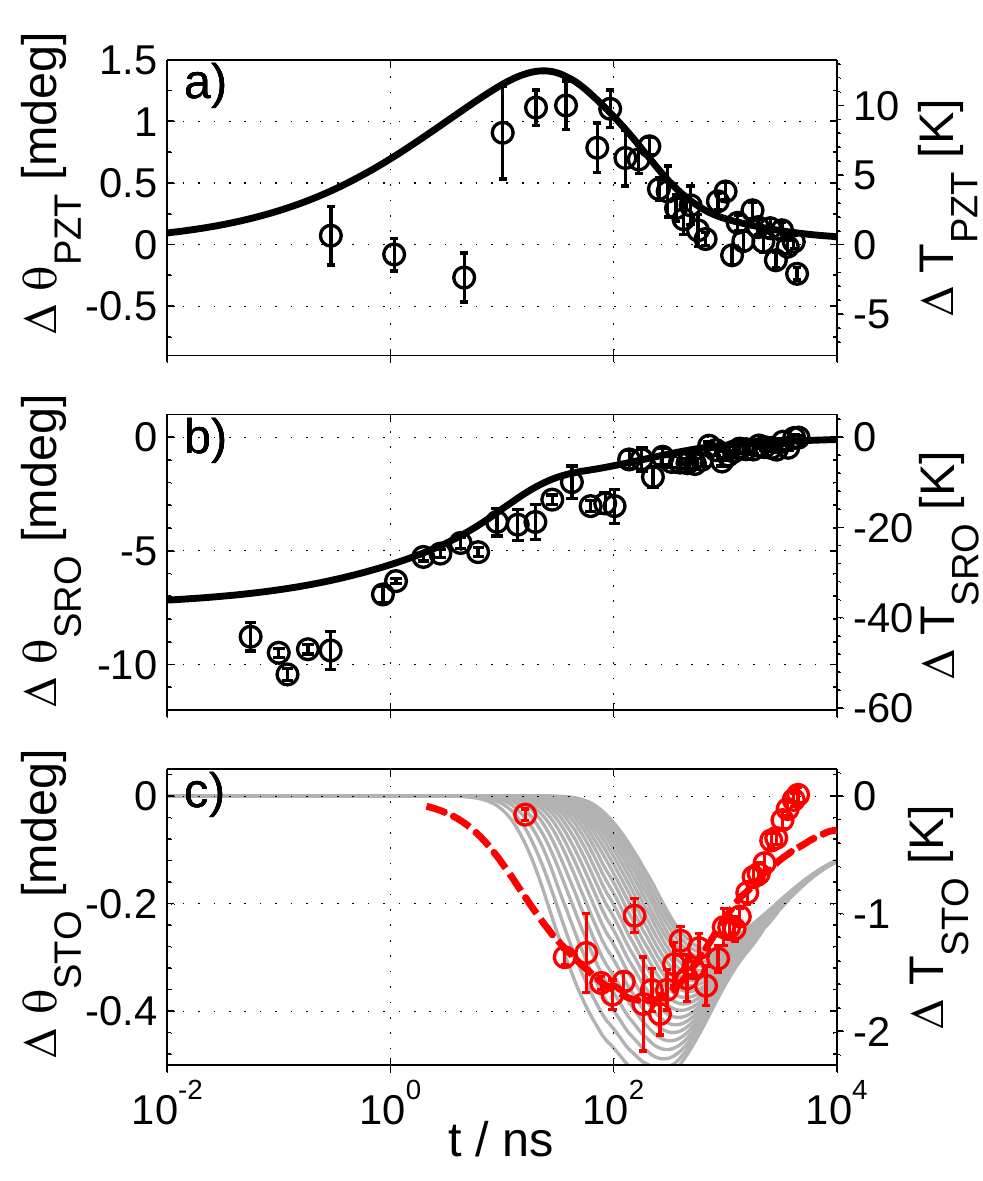}
	\caption{(Color online) Extracted Bragg peak shifts for (a) PZT, (b) SRO, and (c) STO from the measurements (symbols, left vertical axis) and simulations of the average temperatures (solid lines, right vertical axis) for varying time delays. The right vertical axis assumes thermal expansion coefficients valid in thermodynamic equilibrium. In c) the grey lines indicate the temperature for sample depths from $z=10$ to 1000 nm and the dashed red line is a simulation of the Bragg peak shift $\Delta \theta (t)$ using dynamical x-ray diffraction theory. }
	\label{fig:peakshift}
\end{figure}
Based on the derived $\Delta T$and the PZT peak shift (Fig. 1), the average linear thermal expansion coefficient normal to the sample surface of PZT was determined to be approx. $\alpha_\text{PZT} = \SI{-1.1e-5}{K^{-1}}$ for $\Delta T=190$K and $\alpha_\text{PZT} = \SI{-0.66e-5}{K^{-1}}$ for $\Delta T=84$K.
This temperature dependent negative linear thermal expansion of PZT in $z$-direction is consistent with literature values for thin films. It is connected to its ferroelectricity which leads to a structural phase transition of the material from cubic to tetragonal below its Curie temperature of approx. \SI{500}{\celsius}.\cite{jano2007a}
Heating of PZT leads to a nearly negligible volumetric expansion of the material since the $c$-axis contraction is compensating the expansion in the $a$-$b$-plane while approaching the cubic phase.\cite{jano2007a}


For the analysis of the transient rocking curves the Bragg peak shifts $\Delta \theta$ of the three layers after the  laser pulse excitation were extracted by fitting the individual Bragg peaks by Gaussian functions.
The resulting transients $\Delta \theta (t)$ for each material are plotted in Fig.~\ref{fig:peakshift}.
In total, we cover nearly five decades of time scales\cite{navi2012a} without changing the experimental setup, i.e. comparing the diffraction curves for all time scales does not require any intensity scaling and all critical parameters such as the excitation fluence are constant.

The measured peak shift directly yields the strain in each layer via
$\epsilon=-\Delta\theta\ \cot \theta$.\cite{shay2011a}
For some applications this is already the relevant parameter, as the strain is definitely responsible for the fatigue of high-frequency nanodevices. Assuming the literature values for the thermal expansion coefficients of the materials SRO and STO and using the measured value for $\alpha_{PZT}$, we can moreover deduce the temperature rise $\Delta T = -\epsilon / \alpha$  and qualitatively  discuss the observed thermal diffusion and the expected coupled thermoelastic dynamics, keeping in mind that the regular thermal expansion might have to be modified for short timescales if in fact the local population of phonons has strong non-equilibrium character and the thermal stresses are not relaxed yet. For direct reading of temperature changes we nonetheless add a vertical axis to the experimental data in Fig. 2.

\begin{table*}\centering
    \begin{tabular}{r r r@{ } l p{2cm} r@{ } l p{2cm} r@{ } l p{2cm}}\toprule[0.1 em]
                   & & \multicolumn{2}{c}{\textbf{PZT}} &       & \multicolumn{2}{c}{\textbf{SRO}} &      	& \multicolumn{2}{c}{\textbf{STO}} &	\\ \midrule[0.05 em]
         \emph{lin. therm. exp. coefficient} & [1/K]:  	& $-0.55$ & $\times10^{-5}$& (fit/exp.) 	& $0.88$ & $\times10^{-5}$  &(Ref. \citenum{Yama2004}) 	& $1$ & $\times10^{-5}$ & (Ref. \citenum{deli1996a}) \\
        \emph{heat capacity} & [J/mol K]:    	& 120 & &  (Ref. \citenum{ross2005a})      & 114 & & (Ref. \citenum{Yama2004})	& 128 &  &(Ref. \citenum{deli1996a})         			 \\
         \emph{thermal conductivity} &  [W/m K]: & 	3 & & (fit)                			& 1 & & (exp.)             						 & 10 & & (Ref. \citenum{wang2010a})            		 \\
        \bottomrule[0.1 em]
    \end{tabular}
    \caption{Parameters for the simulation of the thermoelastic dynamics with according reference.}
    \label{tab:parameter}
\end{table*}

The optical femtosecond pump-pulses are exclusively absorbed in the metallic SRO layer, thus rapidly heating this material. 
The initial expansion dynamics in SRO is beyond the time-resolution of 100 ps, and the associated strain waves were discussed previously.\cite{schi2013a,schi2013b}
In this paper we focus on the nanosecond time scale.
The SRO peak shift $\Delta \theta_{SRO}=7$mdeg at 1 ns observed in Fig. 2 a) implies a transient temperature rise of $\Delta T_{SRO}=27$K averaged over the SRO layer. 
Fig. 2 b) and c) confirm that at 1 ns the temperature of the adjacent STO and PZT still corresponds to $T = 293$K$+84$K$=377$K, elevated from room temperature according to the average thermal heating of the entire structure by the 208 kHz excitation seen in static experiments. Heat diffusion from the hot SRO to the adjacent materials leads to the cooling of SRO and a temperature increase in the PZT layer and STO substrate. In fact we can quantify their average heating after 100 ns to be $\Delta T_{PZT}=6.6$ K and $\Delta T_{STO}=1.5$K. Surprisingly, the heat essentially flows into the PZT layer, although its thermal conductivity is three times lower than that of the substrate (Table \ref{tab:parameter}). Moreover at this time the temperature of the PZT layer is higher than the initially excited metal layer: $\Delta T_{PZT}(100)$ns $>\Delta T_{SRO}$(100 ns)$= 5.7$K. This is a consequence of cooling the sample via the subtrate with essentially zero heat dissipation at the sample surface. This behavior is emphasized by the exponentially decaying temperature profile in the 147 nm thick SRO layer which essentially heats the material near the PZT interface.
After about 2 $\mu$s the data show that the temperatures in the PZT and SRO layers have equilibrated with the substrate temperature ($\Delta T = 0.1$ K) in the first micrometer, which is sensed by the x-rays. Finally, the transient heat load diffuses deeper into the STO substrate, where the extinction depth of the X-rays makes the thermal strains invisible to the current experiment.

While for delay times larger than 50 ns the measured out-of-plane strain can be consistently interpreted as thermal expansion, the situation is more complex for the earlier dynamics. Although the PZT lattice must undoubtedly be heated between 1 - 10 ns since the temperature of SRO is decreasing by about 30$\%$, Fig. 2a) clearly shows no contraction of PZT during this time.
In thermodynamic equilibrium the out-of-plane contraction of PZT goes along with an in plane expansion.
For our experimental conditions a rough estimation of the lower limit of the time it takes for the complete evolution of in-plain strains can be calculated from the probe diameter $d_\text{Probe} = \SI{50}{\micro m}$ and the sound velocity in SRO. The resulting travel time is approx. $t_{\parallel} = \SI{8}{ns}$ and gives a lower limit of the lateral strains to evolve within the complete region of observation.
This suggests that the epitaxially hindered in-plane expansion prevents the out-of plane contraction to occur.

\begin{figure}
	\centering
	\includegraphics [width=0.8\columnwidth]{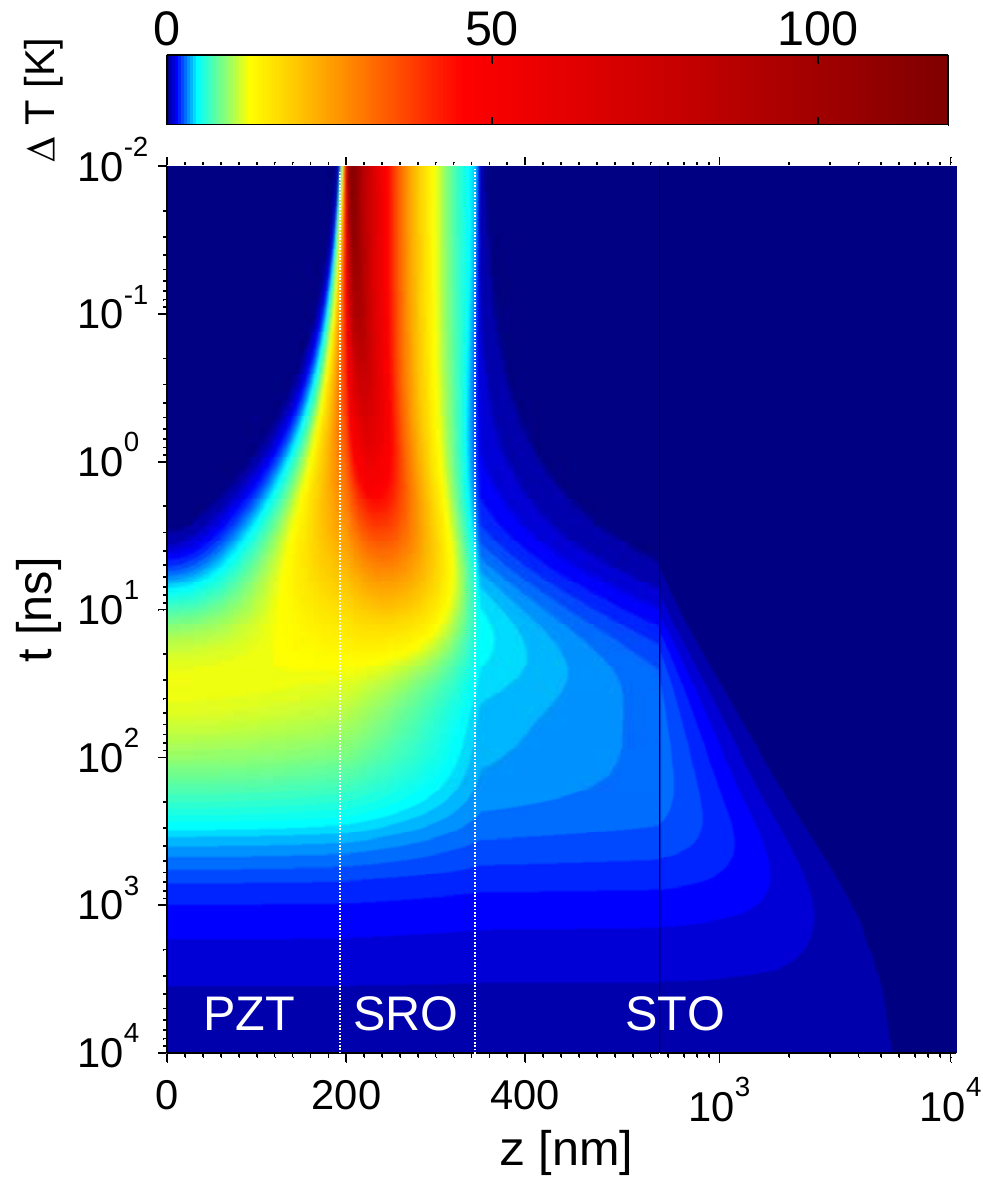}
	\caption{(Color online) The 2D plot shows the results of the 1D heat equation over delay $t$ and depth $z$.
	The color code is not linear.
	Parts of the $x$-scale are logarithmic for the STO substrate.}
	\label{fig:tempMap}
\end{figure}

\textbf{Simulation \& Discussion}

In order to confirm this interpretation, we compare the data with a simplified 1D heat diffusion simulation.\cite{shay2011a}
We calculate the laser-induced temperature jump in the metallic SRO layer from the material's optical penetration depth of 47 nm, its heat capacity and the laser fluence.
The thermal diffusion is modeled by the 1D heat equation.\cite{shay2011a} The result is shown in Fig.~\ref{fig:tempMap} as a color coded temperature on logarithmic time scale, as well as a logarithmic length scale for parts of the STO substrate. These simulations confirm the interpretations of the experimental data given above, in particular the inversion of the temperature rise in PZT and SRO. The exponentially decaying temperature profile in SRO leads to a very large temperature gradient towards PZT and to a weak gradient with respect to STO. This drives the predominant heat flow into PZT at times $t<10 $ns. Around $t=50$ ns the temperature in PZT and SRO is nearly equilibrated and the only direction for heat dissipation is into the substrate.

For verification and analysis of  our interpretation it is sufficient to extract the average heating of SRO and PZT from the heat diffusion simulation in Fig. 3 and plot it together with the measured heat expansion in Fig. 2, using the equilibrium expansion coefficients without additional scaling. The excellent agreement for $t>10$ns show the self-consistency of our analysis. On the interesting timescale $1-10$ns the simulation confirms that PZT is substantially heated after 5 ns, although the out-of plane strain is unchanged.


The analysis of the substrate peak is more complex.  To highlight the problem, we compare the temperatures (grey lines in Fig.~\ref{fig:peakshift}c) obtained in the simulation for different depths of the substrate to the measured peak shift using the bulk expansion coefficient of STO. Alternatively, we use the simulated strain profile to calculate the Bragg peak in the framework of dynamical diffraction and fit the peak position with the same procedure as the experimental data. The result is the red line in Fig. 2 c) which shows excellent agreement up to $t =1 \mu$s. Such analysis, especially of the wings of the substrate peak, was discussed for the case of Si crystals.\cite{lars1983a} At longer time-scales
the heat has diffused into the crystal on the order of the pump-spot size the problem becomes 3D, explaining the deviation. 

The structural parameters for the simulations were all determined by experiments\cite{schi2013a}, and the thermoelastic parameters were taken from literature, see Tab.~\ref{tab:parameter}.
Since all literature values for the simulations were measured statically and for bulk samples, the fit procedure\footnote{The main fit parameters were the absorbed fluence ($F = \SI{1.8}{mJ\per cm\squared}$) of the pump pulses as well as the thermal conductivity of PZT. The resulting $\kappa_{PZT}=3$W/mK agrees well with the range given in the literature. Although we have determined the linear thermal expansion coefficient of PZT as $\SI{-0.66e-5}{K^{-1}}$ in the static measurements, we obtain slightly better agreement for a smaller value of $\SI{-0.55e-5}{K^{-1}}$ to get the best fit for the transient experimental data.
Experiments of our group on the heat diffusion of single SRO thin films on a STO substrate reveal a smaller thermal conductivity of SRO $k_\text{SRO} = \SI{1}{W\per m K}$ in contrast to the literature value of $\SI{5.9}{W\per m K}$ for bulk SRO.\cite{Yama2004}} of the simulation started with the STO substrate Bragg peak shift for late times ($t > \SI{10}{ns}$), see Fig.~\ref{fig:peakshift}~c).

For a direct comparison of the heat diffusion model to the data on the shortest timescales we could calculate the transient thermal stresses from the spatio-temporal temperature map and feed them into a 1D linear chain model (LCM) of masses and springs in order to account for the initial ultrafast phonon-dynamics.\cite{herz2012b, herz2012a} For the 1-10 ns timescale the restriction to one dimension would, however, still not allow for dissipation of in-plane stresses and we would require a dynamical 2D model.

\textbf{Conclusion}

We applied synchrotron-based UXRD experiments to simultaneously follow the out-of-plain thermal strains of three different materials in a thin film sample over nearly five decades of time scales. Material constants are either taken from literature or their measurement is inherent to the present experiment. The data  show that the expitaxial connection to the substrate delays the out-of plane contraction of PZT by about 10 ns, because its in-plane expansion can only take place on the timescale of sound propagation through the excited area.
Moreover the data evidences a transient inversion of the temperature gradient between the excited SRO and the initially cold PZT.
Simple 1D simulations of the temperature profile confirm this feature and show excellent agreement with the data on intermediate times \SI{10}{ns} - \SI{1}{\micro\second} in which the 1D approximation is valid. 
We believe that similar experiments have to be utilized as a reference for according thermoelastic models on the ultrafast time and nanometer length scales, because of the complexity of the included dynamics.

We thank Ionela Vrejoiu for providing the sample,
and the BMBF for the funding via 05K2012 - OXIDE.


\end{document}